\documentclass[twocolumn,prd,showpacs,showkeys]{revtex4-1}
\usepackage{amsmath,amssymb,pgf,graphicx,color,url,array}
\usepackage{tabularx}
\usepackage{wasysym}
\usepackage{longtable}
\newcommand{\be}{\begin{equation}}
\newcommand{\ee}{\end{equation}}

\begin{document}
\begin{flushright}
INR-TH-2018-030
\end{flushright}

\vskip -0.9cm
\title{Spatial structure of the WMAP-Planck haze}

\author{Grigory I. Rubtsov}
\affiliation{Institute for Nuclear Research of the Russian Academy of Sciences, 117312, Moscow, Russia}
\author{Yana V. Zhezher}
\email{zhezher.yana@physics.msu.ru}
\affiliation{Institute for Nuclear Research of the Russian Academy of Sciences, 117312, Moscow, Russia}
\affiliation{Faculty of Physics, M.V. Lomonosov Moscow State University, 119991, Moscow, Russia}

\begin{abstract}
It was proposed that the two phenomena, WMAP-Planck haze and
Fermi bubbles, may have a common origin. In the present paper we
analyze the spatial structure of the haze using the Planck 2018 data
release. It is found that the spatial dimensions and locations of
WMAP-Planck haze and Fermi bubbles are compatible within the
experimental uncertainties. No substructures similar to the Fermi
bubbles cocoon are identified in the Planck data. Comparison with the
spatial extent of possible synchrotron emission caused by the
electron-positron pair emitted by the Galactic center pulsar
population and by the decay of dark matter particles in
the Galactic center region are performed. Both galactic pulsars and
dark matter decay remain viable explanations of the WMAP-Planck
haze.
\end{abstract}
\keywords{WMAP-Planck haze -- Fermi bubbles }
\maketitle

\section{Introduction}
\label{sec:intro}

A phenomenon called the ``WMAP haze'' was found in 2003 by
D.~P.~Finkbeiner by analyzing the maps of the microwave radiation
measured by the Wilkinson Microwave Anisotropy Probe (WMAP)
spacecraft~\cite{Finkbeiner2003}. The experiment aimed at the
precision study of the cosmic microwave background (CMB) anisotropy
measured the full-sky maps in five different frequency bands: 23, 33, 41,
61 and 94 GHz~\cite{WMAP}. Being the images of the entire sky in a
corresponding frequency band, they represent a total radiation
generated by all possible mechanisms. A full-sky map may be
interpreted a sum of a CMB radiation map with an interstellar matter
emission maps and a map of point sources.

In the aforementioned analysis three emission
mechanisms are assumed, which are believed to contribute the most at
the WMAP frequency range: free-free, synchrotron and thermal dust
emission. The contribution of the known point sources and the Galactic
plane is masked out, while the map of CMB radiation is subtracted. In
the Galactic center (GC) region an excess radiation is observed which
may not be attributed to any of the emission mechanisms discussed.  It
appeared to be radially-symmetric in $l$ and $b$ coordinates in both hemispheres, extending up to $20^{\circ}$
outside the Galactic plane, with a hard spectrum $\nu F_{\nu} \propto
\nu^{0.1}$. The existence of the WMAP haze was confirmed by the Planck
mission~\cite{Ade}, and since that the ``WMAP-Planck haze'' has become
a common name for this phenomenon.

A number of theoretical explanations for the WMAP-Planck haze have
been put forward. One may roughly divide all models into three
large categories:

\begin{itemize}
\item First group of theories implies dark matter (DM) explanation for
  the WMAP-Planck excess. Dark matter particles in form of weakly
  interactive massive particles (WIMP) may decay or annihilate and
  produce secondary particles. The latter in turn emit radiation
  through inverse Compton scattering with background photon fields, and through synchrotron emission. In a notable subclass of
  this group, DM is composed of supersymmetric
  particles~\cite{Hooper,Caceres,Linden:2010eu}.

\item Theories related to the second category suggest pulsars as a
  source of WMAP-Planck haze. Pulsars being rapidly rotating neutron
  stars may be a source of electron-positron pairs, which in its turn
  produce an excess of radiation by the very same mechanism mentioned above. The result is based on the
  distribution of pulsars in the inner part of Milky Way~\cite{Kaplinghat,Harding}.

\item A third class of theories suggests a common origin for two or
  more different types of excesses of radiation observed in the
  Galactic Center region: Fermi bubbles~\cite{Su} (hereafter FB), soft
  X-ray excess observed by ROSAT and Suzaku~\cite{Su,Kataoka}, bipolar
  structure found using MSX data at mid-infrared
  wavelengths~\cite{Bland} and 2.3 GHz excess observed by Parkes
  telescope~\cite{Carretti}. According to this scenario the electron
  populations in the Galactic Center regions originate from explosive
  outburst from the central supermassive black hole, see~\cite{Crocker,Dobler} and references therein.
\end{itemize}

With increasing amount of observations, the phenomena described in the
third hypothesis have been constantly studied. High-accuracy results
available now allow one to test third class of theories using
experimental data. The main goal of this paper is a comparison of
spatial structures of the WMAP-Planck haze and the Fermi Bubbles in
order to test the hypothesis of the common origin. For that purpose,
the latest throughout analysis of the Fermi LAT collaboration was used, in which a number of features have been determined~\cite{Ackermann}. These features together with general properties such as extended emission shape and asymmetry in northern and southern hemispheres were investigated in the haze. 

The latest studies~\cite{Pshirkov,Paolis} have shown an indication on the existence of the Fermi bubbles and Planck haze around the M31 galaxy, and it is possible for this phenomena to be a common property for some types of galaxies.

The paper is organized as follows: in Section~\ref{sec:methods} we
describe the template fitting method for deriving the WMAP-Planck haze
map from the Planck 2018 release data. In Section~\ref{sec:data} the
data set and employed emission mechanisms are encompassed, and,
finally, in Section~\ref{sec:results} the results of testing different
scenarios for the WMAP-Planck haze origin are presented. The results
are discussed in Section~\ref{sec:discussion}.

\section{Methods}
\label{sec:methods}

\subsection{Template fitting}
\label{subsec:templatefit}

We follow the template fitting procedure, the method used in the
original analysis~\cite{Finkbeiner2003}, which allows to estimate the
WMAP-Planck haze based on the observed maps. It requires one to
produce templates, full-sky maps created assuming a theoretical
description of specific interstellar matter emission mechanism. The
observed map is fit with $\chi^2$ method as a sum of CMB map and
template maps, weighted with numerical factors, called amplitudes. The
target emission is obtained as a residual of the fit.

Specifically, the Planck map of the microwave radiation is represented as:
\begin{equation}
d_i = a_{\alpha}P_{\alpha i} + x_i,
\end{equation}

\noindent where $\vec{d}$ -- the experimental full-sky map, $P$ -- a
spatial templates matrix, $\vec{a}$ -- an amplitude vector, $\vec{x}$
-- a haze component to be derived. $i$ is the number of pixels in the
map and $\alpha$ determines the number of templates used: three
templates for free-free, synchrotron, dust emission and a specific
template for WMAP-Planck haze. Following~\cite{Finkbeiner2003} the
latter is used at the fit stage to avoid compensation of the haze by
other components.

The statistical uncertainty of the Planck full-sky map may be
represented as a white noise. The latter is considered statistically
uncorrelated between pixels and hence its covariation matrix is
diagonal in the pixel space:

\begin{equation}
\langle d_i d_j \rangle = N_{\underline{i}} {\delta}_{\underline{i} j},
\end{equation}

\noindent where $N_i$ is a vector with length equal to the number of pixels in the map.

The chi-squared criteria can be written as:
\begin{equation}
{\chi}^2 = \sum_i \frac{x_i x_i}{N_i} =\sum_i \frac{{\left(d_i - \sum_{\alpha} a_{\alpha} P_{\alpha i} \right)}^2}{N_i}.
\end{equation}

The optimization of the $\chi^2$ may be done analytically with
differentiation by $a_{\alpha}$, which gives the desired solution for the amplitude vector:

\begin{equation}
\sum_i \frac{d_i P_{\alpha i}}{N_i} = \sum_{i,\alpha '} a_{\alpha '} \frac{P_{\alpha ' i} P_{\alpha i}}{N_i},
\end{equation}

\noindent or, using matrix notation:
\begin{equation}
a_{\alpha '} = {\left( M^{-1} \right)}_{\alpha \alpha'} F_{\alpha}\,,
\end{equation}
where
\begin{equation}
F_{\alpha} = \sum_i \frac{d_i P_{\alpha i}}{N_i},
\end{equation}
\begin{equation}
M_{\alpha \alpha'} = \sum_i \frac{P_{\alpha ' i} P_{\alpha i}}{N_i}.
\end{equation}

\subsection{Haze template}
\label{subsec:hazetemplate}

To prevent other components from totally or partially compensating the
WMAP-Planck haze, a template for the latter is also included into
template matrix at the fit stage. Following~\cite{Ade} and~\cite{Dobler}, we have adopted a two-dimensional Gaussian ellipse with the characteristic lengths $\sigma_l = {15}^{\circ}$ and $\sigma_b = {20}^{\circ}$.

\subsection{Mask}
\label{subsec:mask}

\begin{figure}[h]
\includegraphics[width=0.48\textwidth]{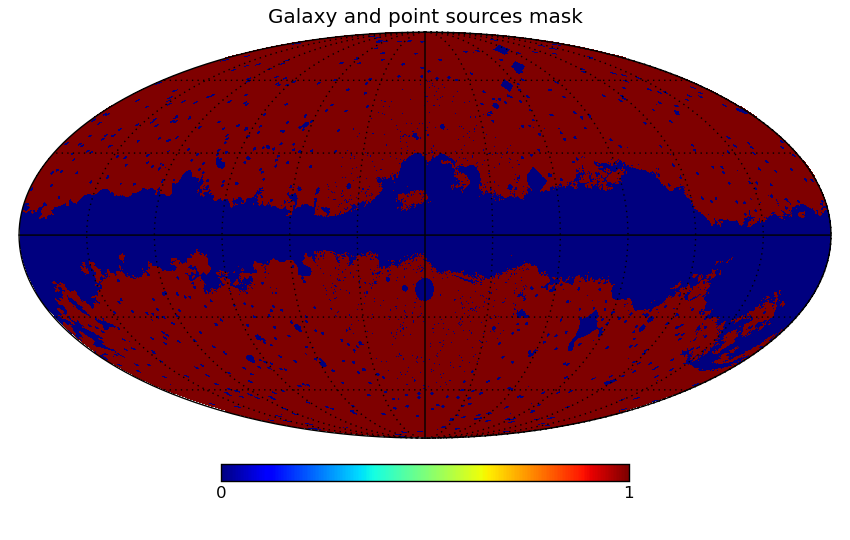}
\caption{The mask used for the present analysis: SMICA mask where areas with the galactic extinction is greater than $1\ \mbox{mag}$ and areas with $H_{\alpha}$ grater than $10\ \mbox{rayleigh}$ are masked. Final mask excludes $30.8\ \%$ of the sky.}
\label{figure:mask}
\end{figure}

All the bright areas of the sky, such as point sources and the Galaxy should be covered with a mask. We have started from the SMICA mask and then have masked areas where unknown systematics dominates and which may include unresolved point sources. To do that, we went beyond the cuts suggested by~\cite{Ade} and covered areas with the galactic extinction greater than $1\ \mbox{mag}$ and areas with $H_{\alpha}$ greater than $10\ \mbox{Rayleigh}$. Final mask excludes $30.8\ \%$ of the sky and is shown on Fig.~\ref{figure:mask}.

\section{Data set}
\label{sec:data}

\subsection{Planck microwave emission maps}
\label{subsec:initial}

Originally the WMAP-Planck haze was found in the WMAP data, and it appeared to be the brightest at K, Ka and Q bands (23 GHz, 33 GHz and 41 GHz, see~\cite{Finkbeiner2003}).

The Planck satellite was launched in 2009 with the aim for the measurement of the CMB anisotropies with the unprecedented precision~\cite{Tauber}, and it has been operating till 2013.

For the observations at the lower end of the range, the Planck
satellite is equipped with the Low Frequency Instrument
(LFI)~\cite{Davis,Varis}. As as part of the Planck 2018 release the full-sky maps at
three frequency bands: 30 GHz, 44 GHz and 70 GHz are provided~\cite{AdeLFI,PLA}, which
are used as an initial data for the WMAP-Planck haze separation in the
present Paper.

\subsection{ISM emission templates}
\label{subsec:templates}

As was discussed above, one should include emission mechanisms relevant for energies of interest.  For the microwave frequency band, three general emission mechanisms are usually adopted:

\begin{itemize}
\item free-free emission -- emission due to interaction between electrons and protons at the regions of ionized hydrogen ($HII$). $H_{\alpha}$ emission (Lyman line transition of ionized hydrogen) is used as a template for this mechanism. As a free-free template, we have used $H_{\alpha}$-emission map created by Finkbeiner~\cite{Finkbeiner2003a}, based on wide-angle observations of Virginia Tech Spectral line Survey (VTSS), Southern H-Alpha Sky Survey (SHASSA) and Wisconsin H-Alpha Mapper (WHAM).
\item synchrotron emission -- an emission of electromagnetic waves by charged particles which propagate in a magnetic field with relativistic speeds. The template created by Remazeilles et al.~\cite{Remazeilles} was used. It is the reprocession of the ``408 MHz All Sky Continuum Survey'' map created by Haslam in 1982~\cite{Haslam1,Haslam2}. The original map was created with 408 MHz surveys from Jordell Bank MkI telescope and Effelsberg 100 meter telescope. Additionally, to cover the whole sky, two additional surveys were made by Parkes 64 metre telescope and Jordell Bank MkIA telescope. In 2015, Remazeilles et al. have improved the initial 408 MHz survey by removing extragalactic sources and reducing large-scale striations.
\item dust emission -- emission of microscopic particles, which predominantly consist of silicon and carbon. Dust particles have an important property which allows to detect it, which is an ability to absorb and scatter light, called extinction. Two different components contribute to what is referred as dust emission: thermal dust and spinning dust with different emission features and properties. $100 \mu m$-map of sub-millimeter and microwave emission of diffuse interstellar thermal dust in the Galaxy was built using COBE/DIRBE and IRAS/ISSA data~\cite{Schlegel}. The spinning dust template is modelled by Planck collaboration using COMMANDER~\cite{Adea}.
\end{itemize}

\section{Results}
\label{sec:results}

\begin{figure}
\includegraphics[width=0.95\columnwidth]{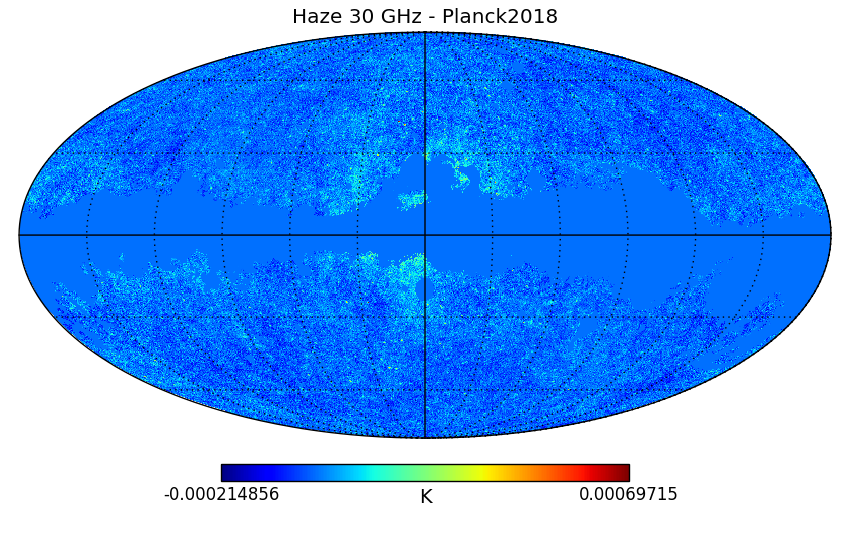}
\caption{The WMAP-Planck haze map obtained for 30GHz Planck 2018 LFI data.}
\label{figure:30GHz}
\end{figure}

\begin{figure}
\includegraphics[width=0.95\columnwidth]{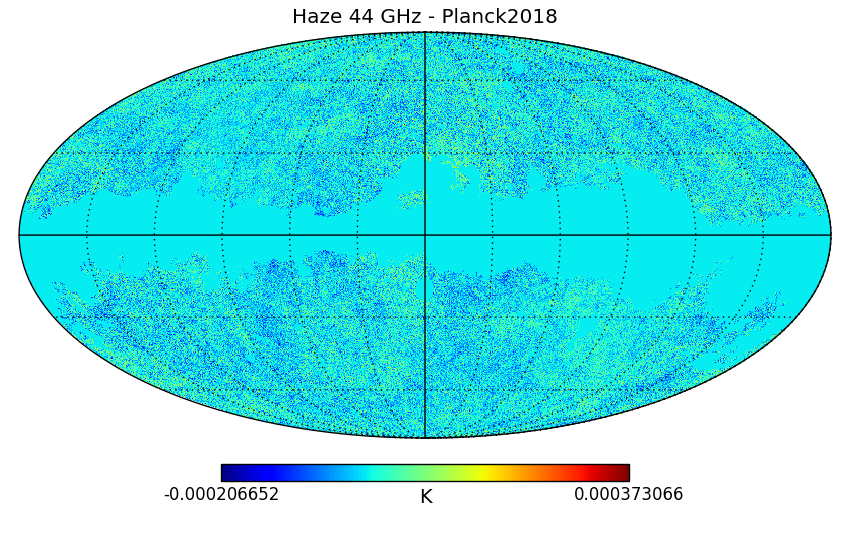}
\caption{The WMAP-Planck haze map obtained for 44GHz Planck 2018 LFI data.}
\label{figure:44GHz}
\end{figure}

The WMAP-Planck haze was derived for Planck LFI maps at 30 and 44 GHz~(see Fig.~\ref{figure:30GHz} and Fig.~\ref{figure:44GHz}). As was expected from the initial analysis~\cite{Finkbeiner2003}, the use of 70 GHz map does not result in any trace of ``haze'' in the GC region. The WMAP-Planck haze appeared to be the most discernible for the 30 GHz frequency band, and this emission map was used for the spatial structure analysis.

\subsection{Spatial dimensions of the WMAP-Planck haze}
\label{subsec:resultssizes}

The spatial extent of the WMAP-Planck haze was approximated with two
ellipses which are defined with three free parameters: ${\sigma}_b$,
${\sigma}_l$ and $b_0$, where ${\sigma}_b$ and ${\sigma}_l$ are
characteristic lengths of an ellipse and $b_0$ is the position of the
center of an ellipse. The result of minimization in case, when the two
ellipses have a common point at the Galactic Center, ${\sigma}_b =
b_0$ is:

\begin{equation*}
{\sigma}_l = {38.7}^{\circ} \pm {5.4}^{\circ}, {\sigma}_b = b_0 = {20.0}^{\circ} \pm {3.7}^{\circ}.\
\end{equation*}

\subsection{North-south asymmetry}
\label{subsubsec:asymmetry}

Asymmetry between the WMAP-Planck haze in the Northern and the Southern hemispheres is analyzed. For this purpose, we have calculated the average emission intensity for both parts in the region constrained by the haze spatial sizes obtained in Section~\ref{subsec:resultssizes}:

\begin{eqnarray*}
<I>_{\mbox{North}} = \left( 4.0 \pm 0.1 \right) \times {10}^{-5} \ \mbox{K},\\
<I>_{\mbox{South}} = \left( 3.2 \pm 0.5 \right) \times {10}^{-5}\ \mbox{K}.\\
\end{eqnarray*}

The derived emission appeared to be more intensive in the Northern
hemisphere. Let us note, however, that similar to the analysis of~\cite{Ade}
the mask is more restrictive in the Northern part of the haze and one
may not exclude contribution of the galactic plane emission which is
not covered by the mask.

\subsection{Spectrum}
\label{subsubsec:spectrum}

\begin{figure}[h]
\includegraphics[width=0.48\textwidth]{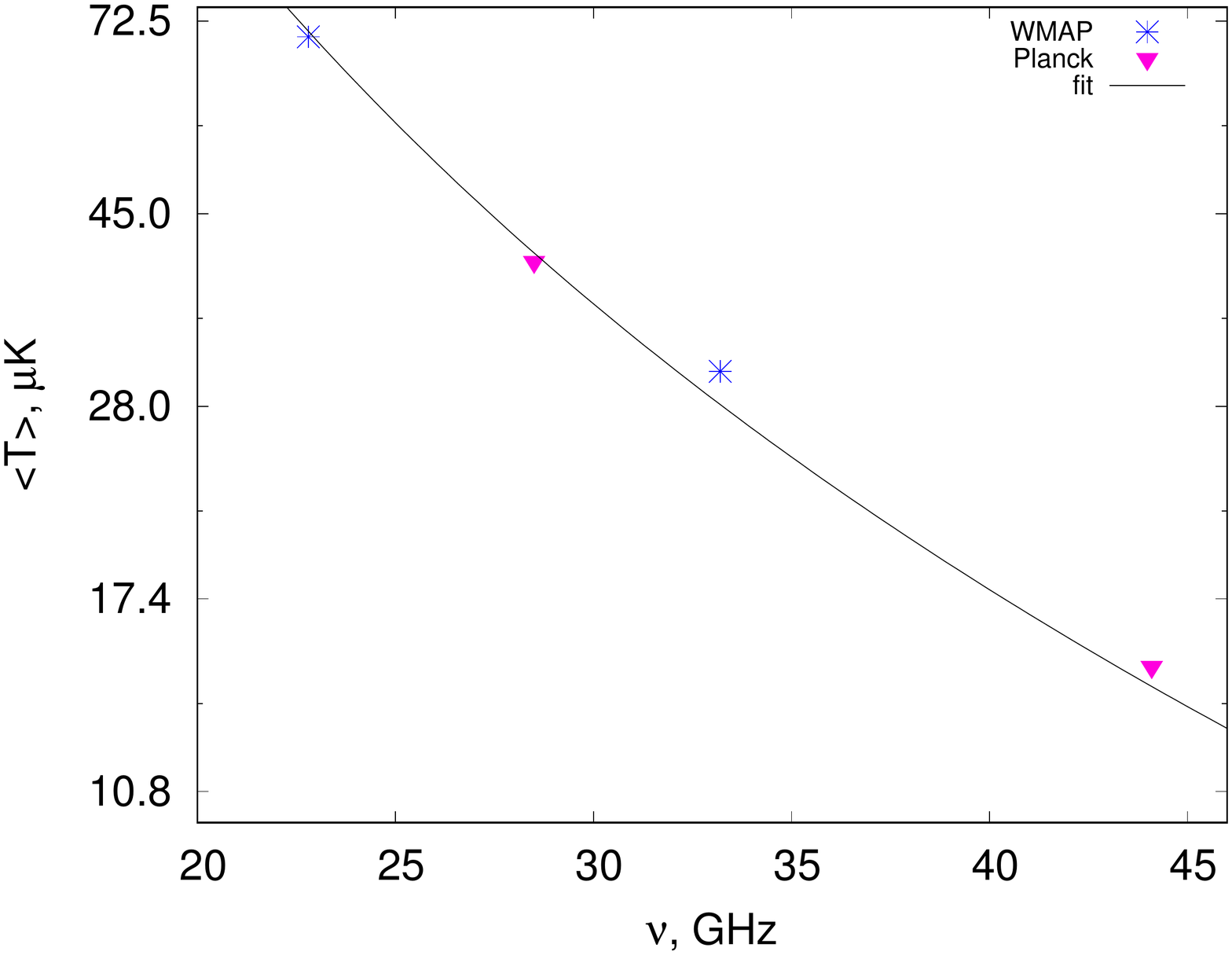}
\caption{The WMAP-Planck haze spectrum: initial data is represented with red crosses, black solid line corresponds to fit for $T_\nu \propto \nu^{\beta}$. Derived result for parameter: $\beta  = -2.45 \pm 0.15$.}
\label{figure:spectrum}
\end{figure}

One of the important properties of the WMAP-Planck haze is its
spectrum, as it may characterize the origin of this phenomena. In
order to consider more spectral points, the analysis of the present
paper is repeated with the nine-year WMAP data in K and Ka frequency
ranges~\cite{Bennett:2012zja,LAMBDA}. The spectrum of the haze
amplitude is shown in Fig.~\ref{figure:spectrum}. The fit of the
spectrum with a power law $T_\nu \propto \nu^{\beta}$ results in the
spectrum slope $\beta  = -2.45 \pm 0.15$, which is in a good agreement with the results obtained by the Planck collaboration: $\beta = -2.55 \pm 0.5$~\cite{Ade}.

\subsection{Spatial structure of the WMAP-Planck haze in comparison with Fermi bubbles}
\label{subsec:resultsfb}

Let us move on to the comparison of characteristic features comparison
between the WMAP-Planck haze and the Fermi bubbles.

\begin{figure}[h]
\includegraphics[width=0.48\textwidth]{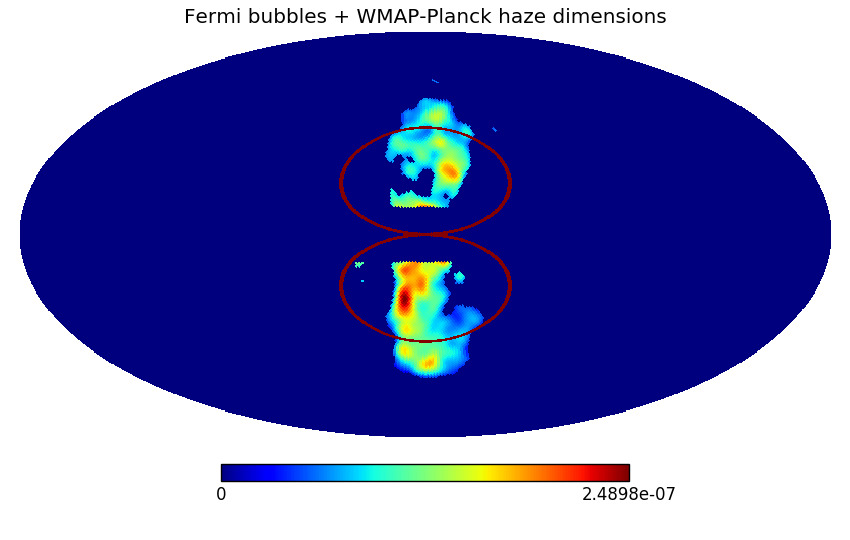}
\caption{The WMAP-Planck haze spatial dimension approximation compared to Fermi Collaboration results (\cite{Ackermann}).}
\label{figure:fermiplushazeassym}
\end{figure}

The size approximation compared to the Fermi collaboration
results~\cite{Ackermann} is shown in the
Figure~\ref{figure:fermiplushazeassym}, which supports the ``common
origin'' scenario on the qualitative level, in which the haze emission
is produced at the reverse shock from the Fermi bubbles material,
tangential to their position. The mechanism for the unified outflow
origin of the haze and other extended emissions at the GC region is in
detail described in~\cite{Crocker}.

\begin{figure}[h]
\includegraphics[width=0.48\textwidth]{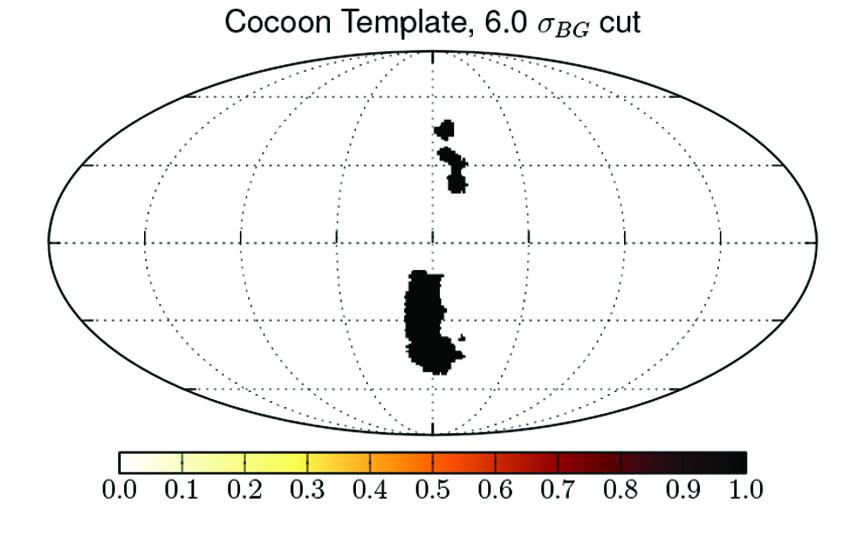}
\caption{The substructures in the Fermi bubbles according to~\cite{Ackermann}.}
\label{figure:cocoons}
\end{figure}

In the latest Fermi bubbles analysis~\cite{Ackermann}, the existence of substructures was claimed both in the northern and in the southern bubbles. These substructures were called ``cocoons'' (see Fig.~\ref{figure:cocoons}). The same assumption was examined for the derived WMAP-Planck emission as well. The northern cocoon appeared to be almost completely covered by the mask, so only the southern one was considered. For this purpose mean intensity over the southern cocoon and over the southern bubble were compared:

\begin{eqnarray*}
<I>_{\mbox{cocoon}} = \left( 2.5 \pm 0.5 \right) \times {10}^{-5}\ \mbox{K},\\
<I>_{\mbox{bubble}} = \left( 3.2 \pm 0.5 \right) \times {10}^{-5}\ \mbox{K}.
\end{eqnarray*}

As follows from the calculated intensities, the substructures in the
WMAP-Planck haze are not observed.

\subsection{Test of the pulsar origin for the WMAP-Planck haze}
\label{subsec:resultspulsars}

\begin{figure}[h]
\includegraphics[width=0.48\textwidth]{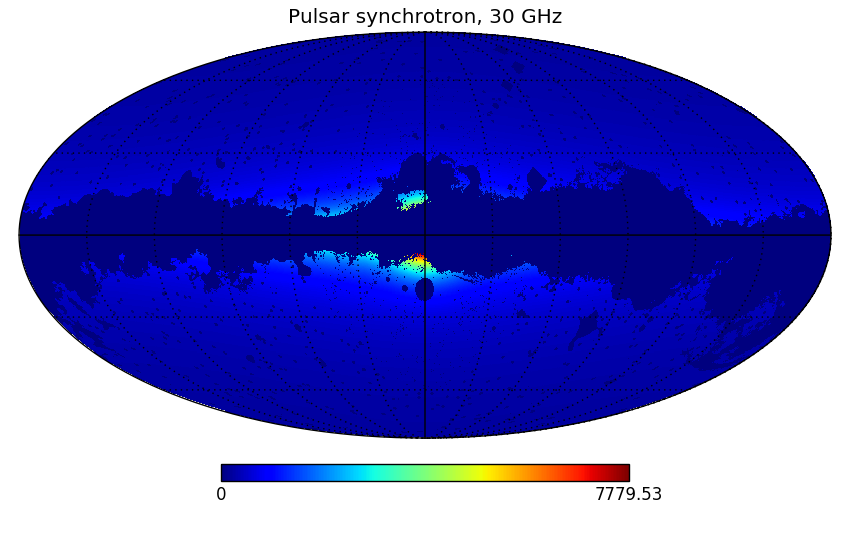}
\caption{Synchrotron flux at 30 GHz from the GC pulsar population based on~\cite{Kaplinghat}, with the mask applied.}
\label{figure:pulsars30GHz}
\end{figure}

Pulsar scenario of the WMAP-Planck haze origin was tested based on the work of M.~Kaplinghat et al.~\cite{Kaplinghat}, where the haze is formed by the synchrotron emission from the electron-positron pairs injected to the interstellar medium by the galactic population of pulsars.

The synchrotron flux was calculated with the use of the GALPROP
propagation package~\cite{Moskalenko:1997gh,Strong:1998pw} according
to the pulsar emission parameters proposed in~\cite{Kaplinghat} which
employ the exponential distribution of sources near the GC, power-law
injection spectrum with the cutoff energy $E_{cut} = 100\ \mbox{GeV}$
and exponential form of galactic magnetic field.

The resulting map for synchrotron flux from the galactic pulsar
population at 30 GHz is shown in Fig.~\ref{figure:pulsars30GHz}.

First of all, spatial sizes for the emission in the Galactic Center
region from the population of pulsars at 30 GHz were obtained in the
similar way as in the Section~\ref{subsec:resultssizes}. The obtained results are:

\begin{equation*}
{\sigma}_l^{p} = {55.15}^{\circ} \pm {0.1}^{\circ}, {\sigma}_b^{p} = b_0^{p} = {9.10}^{\circ} \pm {0.05}^{\circ}.
\end{equation*}

One may go further and apply the template fitting procedure described in
Section~\ref{subsec:templatefit} to the initial Planck 30~GHz map,
adding the pulsar emission map as one of the templates to check
whether it will compensate for the haze. For the derived residual
``haze'' map we have calculated the mean intensity $<I>_{\mbox{in}}$
in the haze region constrained by the haze size derived in the
Section~\ref{subsec:resultssizes} with the mean intensity
$<I>_{\mbox{whole}}$ of the whole map. The results are the following:

\begin{eqnarray*}
<I>_{\mbox{in}} = \left( 1.1 \pm 0.1  \right) \times {10}^{-5}\ \mbox{K},\\
<I>_{\mbox{whole}} = \left( 1.0 \pm 0.1 \right) \times {10}^{-5}\ \mbox{K},
\end{eqnarray*}

\begin{figure}[h]
\includegraphics[width=0.48\textwidth]{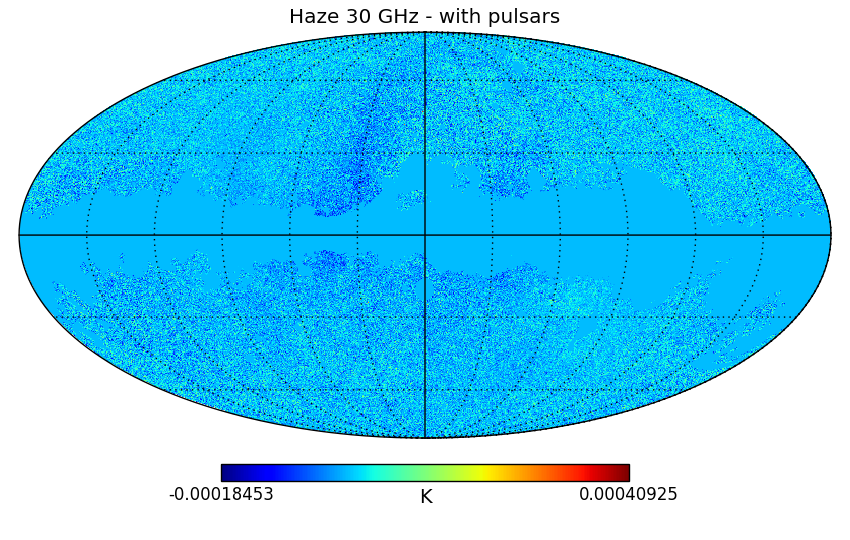}
\caption{Residual map at 30 GHz derived from the template fitting with the pulsar emission map added as one of the templates.}
\label{figure:hazepulsars30GHz}
\end{figure}

\noindent and the corresponding residual map is shown in Figure~\ref{figure:hazepulsars30GHz}.

\subsection{Test of the dark matter origin for the WMAP-Planck haze}
\label{subsec:resultsdm}

\begin{figure}[h]
\includegraphics[width=0.48\textwidth]{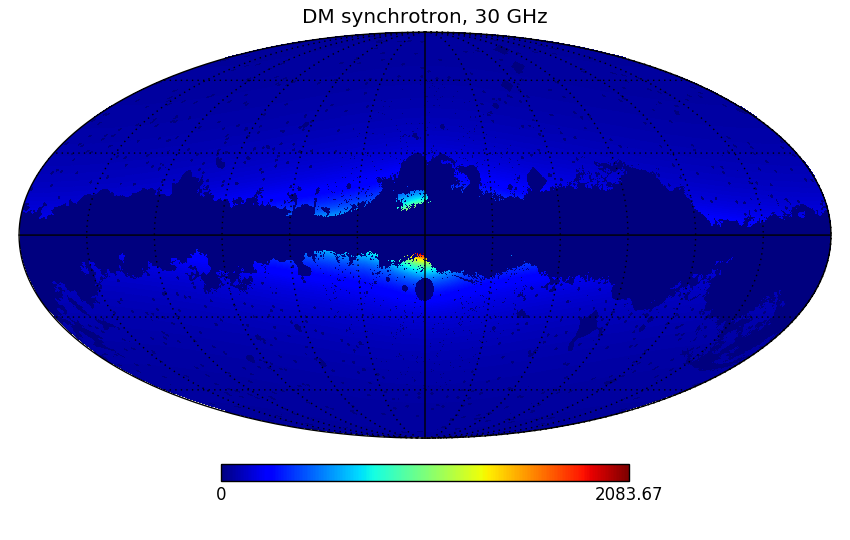}
\caption{Synchrotron flux at 30 GHz from the dark matter decay at the Galactic center region based on~\cite{Linden:2010eu}, with the mask applied.}
\label{figure:DM30GHz}
\end{figure}

Another potential mechanism to generate the WMAP-Planck haze is the synchrotron emission from the electron-positron pairs which occur in the dark matter decay process. The dark matter scenario was tested on the basis of the work~\cite{Linden:2010eu}.

The employed model is the supersymmetric wino dark matter model with mass $400\ \mbox{GeV}$ and with the 100\% decay branching ratio into $W^+ W^{-}$ pairs.

As with the pulsar case, the corresponding synchrotron flux was calculated with the use of the GALPROP package~\cite{Moskalenko:1997gh,Strong:1998pw} with parameters suggested in~\cite{Linden:2010eu}. Namely, we employ the Burkert profile for the dark matter distribution ${f_{Burkert} \left( x \right) = {\left( 1 + x \right)}^{-1} {\left( 1 + x^2 \right)}^{-1}}$, while the electron-positron spectra for the corresponding decay channel are taken from the PPPC4DM~\cite{Cirelli:2010xx}. The derived map for the synchrotron emission at 30 GHz is shown in the Figure~\ref{figure:DM30GHz}.

\begin{figure}[h]
\includegraphics[width=0.48\textwidth]{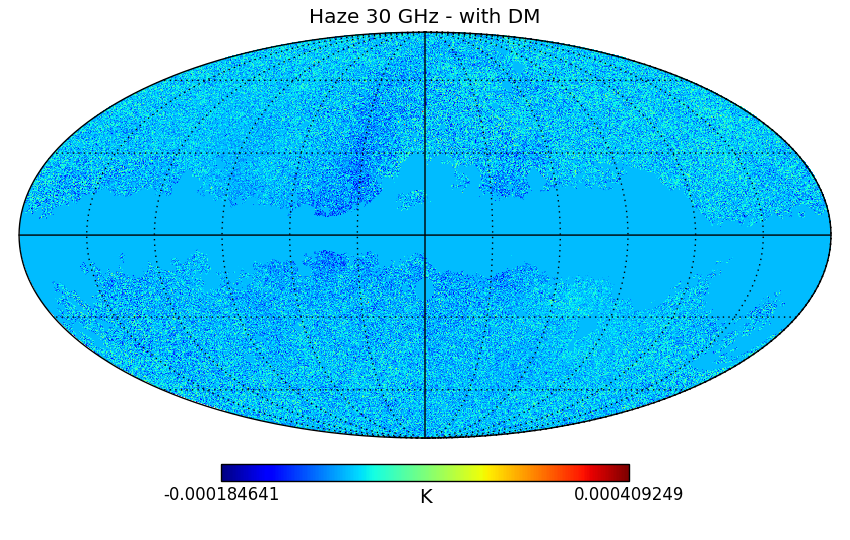}
\caption{Residual map at 30 GHz derived from the template fitting with the DM decay emission map added as one of the templates.}
\label{figure:hazeDM30GHz}
\end{figure}

To test the dark matter origin scenario of the WMAP-Planck haze, the same analysis as for the pulsar case was performed. Template fitting procedure with DM as one of the templates resulted in the residual map shown in the Figure~\ref{figure:hazeDM30GHz}.

Obtained spatial sizes for the synchrotron emission from the propagation of electron-positron pairs from the dark matter decay are:

\begin{equation*}
{\sigma}_l^{DM} = {53.5}^{\circ} \pm {0.2}^{\circ}, {\sigma}_b^{DM} = b_0^{DM} = {9.3}^{\circ} \pm {0.1}^{\circ}.
\end{equation*}

While comparison of mean intensities within the haze region and at the whole sky shows the following results:

\begin{eqnarray*}
<I>_{\mbox{in}} = \left( 1.20 \pm 0.15  \right) \times {10}^{-5}\ \mbox{K},\\
<I>_{\mbox{whole}} = \left( 1.00 \pm 0.15 \right) \times {10}^{-5}\ \mbox{K}.
\end{eqnarray*}

\section{Discussion}
\label{sec:discussion}

The WMAP-Planck haze is derived for the Planck LFI 2018 maps. The
spatial sizes of the haze agree with the one of Fermi bubbles, therefore being in favor of the ``common origin''
scenario. Nevertheless, no substructures similar to the Fermi bubbles
cocoon are identified in the Planck data. Moreover, both the
galactic pulsar and DM emission maps may compensate for the haze
emission if used as an additional template in the template fitting
procedure. Hence, none of the three model classes discussed may be
ruled out as a possible explanation of the WMAP-Planck haze.

\section*{Acknowledgments}
\label{sec:acknowledgments}

The authors are indebted to Maxim Pshirkov and Sergey Troitsky for inspiring and useful discussions. The work is supported by the Russian Science Foundation grant 14-22-00161.


\end{document}